\documentclass[12pt]{article}

\usepackage[english]{babel}

\usepackage[a4paper,top=3cm,bottom=3cm,left=3cm,right=3cm,marginparwidth=1.75cm]{geometry}

\usepackage{amsmath}
\usepackage{graphicx}
\usepackage[colorlinks=true, allcolors=blue]{hyperref}

\title{Analog Time Multiplexing for Digital-to-Analog Conversion}

\author{Juana M. Martínez-Heredia and Alfredo P. Vega-Leal}

\begin{document}
\maketitle

%
\begin{abstract}
The signal bandwidth of Digital to Analog Converters based on Sigma Delta Modulation is limited by speed constrains. Time-Interleaving allows coping with complexity vs. speed by replacing the original architecture by M parallel paths. These path are clocked at a frequency M times smaller  and their   digital outputs time multiplexed. This is then converted to analog by means of a Digital to Analog Converter clocked at the high rate. This preprint proposes that time multiplexing be performed in the analog domain. As a result robustness against dynamic effects is achieved.
\end{abstract}

%
\section{Introduction}
Sigma Delta Modulation (SDM) is receiving increasing attention in the implementation of modern communication standards, both in the design of transmitters and receivers. Wideband  transmitters require the digital-to-analog converter (DAC) to be rated at a very high frequency \cite{li201821,cordeiro2014gigasample,tanio2017fpga}. 

In SDM-based Digital to Analog Converters (DAC)  speed constraints of the technology impose some limitations. If moderate oversampling ratio (OSR) values are chosen, moderate dynamic range (DR) values can be achieved; on the other hand, if a high OSR were used, the signal bandwidth would be low \cite{martin2016multiphase,he2025100,candy1991oversampling}. Time-Interleaving (TI) has been proposed to alleviate speed constraints by trading off complexity and speed  \cite{khoini1997time,pham2008time,colodro1996cellular,kozak2000efficient}. The TI modulator can be derived from the classical one by replicating a basic modulator in $M$ parallel paths clocked at the rate $f_L=1/T_L$, such that the effective sampling rate is $f_H = M f_L =1/TH$. Therefore, the effective OSR is $M$ times greater than that of the basic modulator. Thanks to the current scale of integration, the higher complexity introduced by a TI implementation does not represent a significant problem in terms of cost, power consumption, and silicon area

\begin{figure}[ht]
    \centering
    \includegraphics[width=9cm]{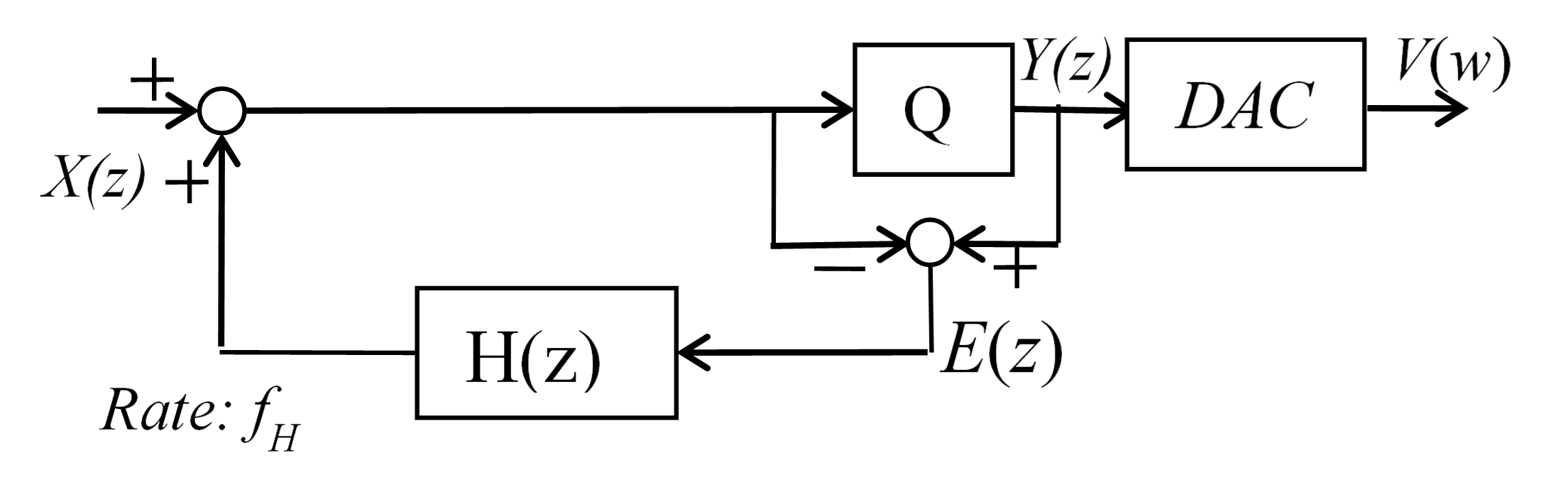}
    \caption{EF-SDM architecture.}
\end{figure}

The Error-Feedback (EF) structure shown in Fig. 1 is commonly used in the digital implementation of SDM based DACs \cite{candy1991oversampling}. In this architecture, clocked at the high rate $f_H$, the quantization error $E(z)$ is fed back to the input through the filter $H(z)$. Thus, the input is directly transferred to the output while the quantization error is affected by the Noise Transfer Function (NTF) \cite{colodro2010spectral},

\begin{equation}
    Y(z) = X(z) + NTF(z) E(z)
\end{equation}

The loop filter is designed so that the NTF is a high pass filter

\begin{equation}
    NTF(z) = 1 + H(z) = (1 - z^{-1})^L/D(z)
\end{equation}

where $L$ is the order of the SDM and $D(z)$ is equal to one for $L = 1$ or $L=2$. If $L>2$, $D(z)$ must be a polynomial of order $L$ that stabilizes the loop \cite{candy1991oversampling,colodro2014linearity}.

\begin{figure}[ht]
    \centering
    \includegraphics[width=11cm]{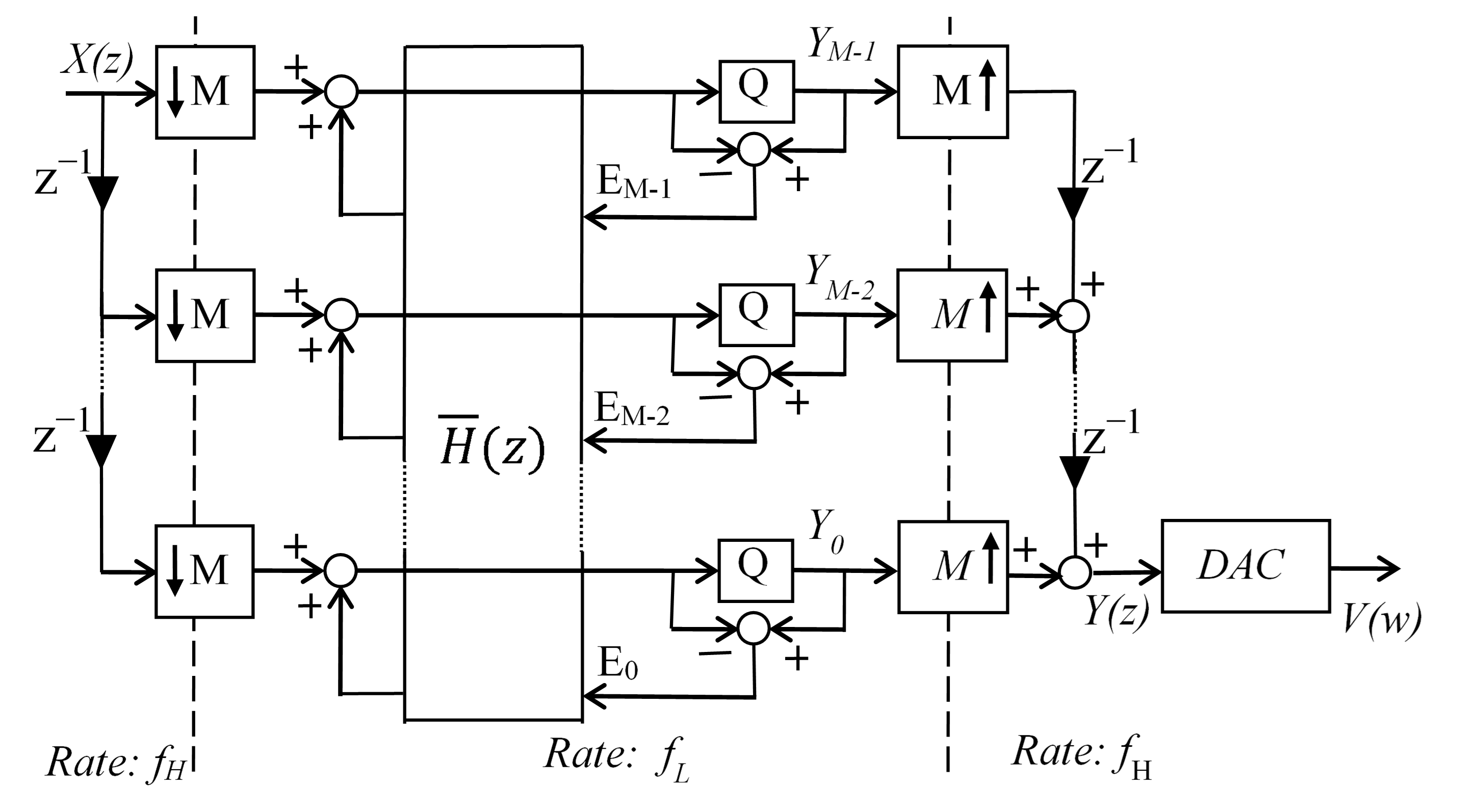}
    \caption{Scheme for the TI SDM based DAC.}
\end{figure}

Using the filtering approach \cite{arahal2016harmonic,pham2008time,kozak2000efficient,colodro2010continuous}, the architecture in Fig. 1 can be converted to its TI equivalent architecture (Fig. 2). The delayed replicas of the input $X(z)$ are separated in $M$ different paths and downsampled to the low rate $f_L$. The filtering is performed at the low rate by means of the block filter $\Bar{H}(z)$. In order to rebuild the high-rate output $Y(z)$, the outputs of the block filters are aligned in time by the up-samplers and the unit delays, and time multiplexed by the adders. The block filter is calculated from the polyphase decomposition terms of $H(z)$ [7, 9], and the output $Y(z)$ in Fig. 2 is a $M$-unit delayed replica of $Y(z)$ in Fig. 1.

The added complexity introduced by the TI SDM based DAC is acceptable when the speed requirements are compromised by technological limitations \cite{arahal2021adaptive,colodro2003multirate}. When the technology prevents the design of the digital SDM at the rate $f_H$, TI is a solution. Unfortunately, the DAC used for converting the digital output $Y(z)$ to its analog counterpart$ V(w)$ becomes the limiting block in Fig. 2, since it still works at the high rate $f_H$. In order to overcome this bottleneck, the time multiplexing of the block filter outputs can be performed in the analog domain, so that the high speed DAC in Fig. 2 can be replaced by $M$ DACs in parallel clocked at the low rate $f_L$. These DACs are activated by $M$ clocks out of phase with each other the amount $2 \pi/M$. These phase shifts implement the unit delays shown in the time multiplexer of Fig. 2. Finally, the DACs’ outputs are summed by means of an analog adder.

Parallel DACs working in TI mode have traditionally been used in Nyquist converters \cite{cheng2011nyquist,colodro2020open}. The increase of the effective conversion rate removes the closest Nyquist images and facilitates the design of the reconstruction filter [16]. Unfortunately, errors in the DACs’ gains and among the parallel analog paths can deteriorate the performances of the system. These errors are produced by mismatch among electronic components. They can be mitigated by a careful layout \cite{deveugele2004parallel,colodro2009analog}, randomization in the use of DACs or digital correction \cite{hovakimyan2019digital}.

TI DACs can also be used in SDM based DACs to relax the speed requirements  \cite{arahal2024multi,mccue2016time}. Unfortunately, unlike the Nyquist converter, the signal at the input of the TI SDM based DAC has a large amount of quantization noise at high frequencies. Non-ideal effects can demodulate this high-frequency noise into the signal band. On the other hand, the inherent filtering of analog multiplexing can achieve greater robustness against timing skew \cite{colodro2022time}.

The architecture of the TI SDM based DAC is described in Section 2. The analysis of how the clock jitter can impact the SNDR of the architectures in Fig. 1 and Fig. 2 is carried out in Section 3. It will be shown there that the TI architecture is more robust against jitter than the conventional DAC.

%
\section{Proposed DAC Concept}

The output stage of the proposed SDM based modulator is depicted in Fig. 3.a. The outputs of the TI SDM are converted to the analog domain through a battery of single-bit and non-return to zero (NRZ) DACs. Unlike the DACs in Fig. 1 and Fig. 2, which are clocked at the high rate $f_H$ (with pulse width $T_H$), the DACs in Fig. 3.a generate their outputs at the low rate $f_L$ using NRZ pulses of width $T_L = 4 T_H$. The conversion cycle starts at the rising edge of the TL -period clocks cp(t) with $p = 0, 1, ..., M-1$. The unit delays shown in Fig. 2 are performed by means of shifting the relative phase of the parallel DACs. If the phase reference is the sampling time of any of the low rate signals $Y_p(z)$, the rising edge of $c_p(t)$ is delayed $p TH$ seconds.

\begin{figure}[ht]
    \centering
    \includegraphics[width=12cm]{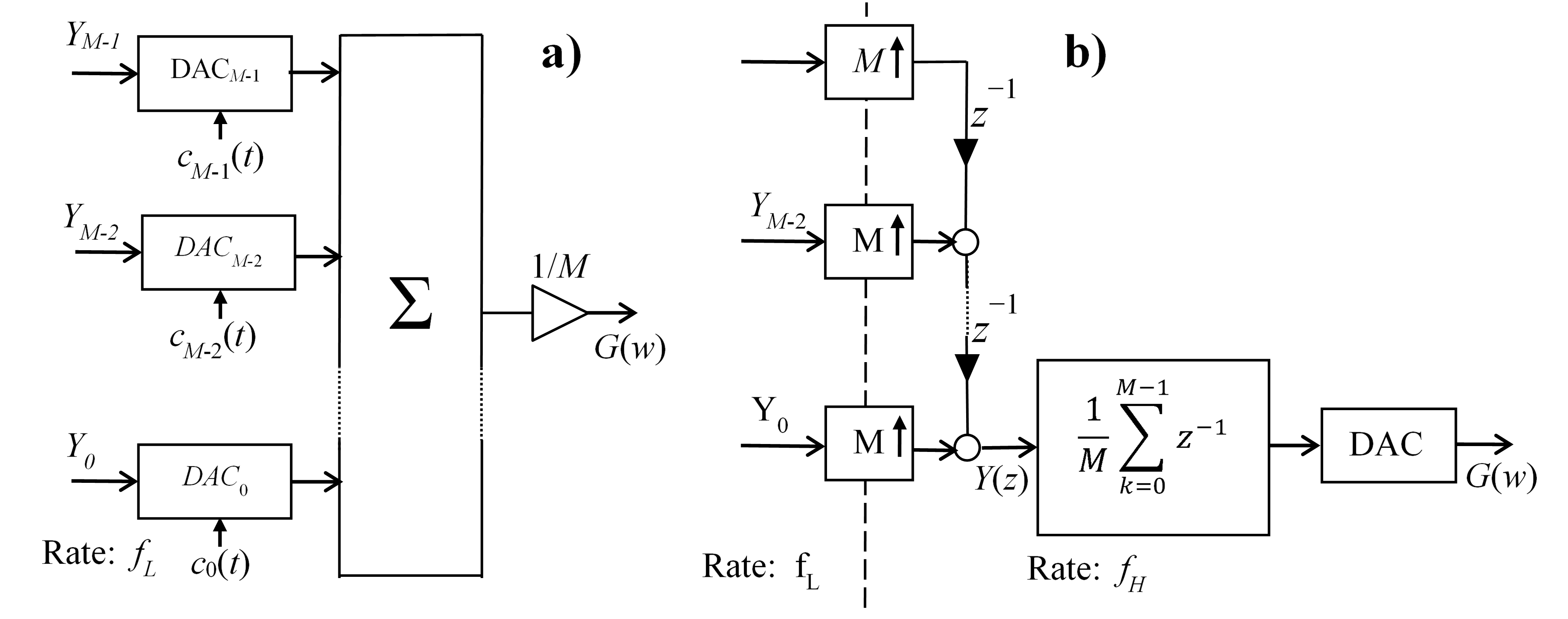}
    \caption{a) Output stage of the SDM based DAC and b) the equivalent discrete-time model.}
\end{figure}

The equivalent discrete-time model
An analysis in the time domain will be carried out to obtain the equivalent discrete-time (DT) model of the circuit in Fig. 3.a. In that follows, the discrete times for the signals at high and low rates will be denoted by the integer variables $n$ and $q$, respectively. The relation between them is $n=qM+r$, where $q=QM(n)$ and $r=RM(n)$ are the integer part and the remainder of the quotient $n/M$, respectively. The sampling times of the signals at high and low rates are $nTH$ and $qTL$, respectively. Finally, $p_k(t)$, with $k=1$ or $M$, form a rectangular NRZ pulse that takes value 1 from $t=0$ to $kT_H$, and 0 the rest of the time.
 
The DAC output in Fig. 2 is
\begin{equation}
v(t) = \sum_{n=-\infty}^{\infty} y(n) p_1 (t-nT_H) 
\end{equation}
\noindent where
\begin{equation}
y(n) = y_r (q)=y_{R_M} (n) Q_M (n)
\end{equation}

For the DAC in Fig.3.a, the output is
\begin{equation}
g(t)=1/M \sum_{q=-\infty}^\infty \sum_{r=0}^{M-1} y_r (q) p_M (t-qT_L-rT_H )
\end{equation}

Now, it is worth noting that
\begin{equation}
p_M (t) =  \sum_{k=0}^{M-1} p_1 (t-kT_H )
\end{equation}

From (5) and (6) it is possible to arrive at
\begin{equation}
g(t)=1/M \sum_{k=0}^{M-1} \sum_{q=-\infty}^\infty \sum_{r=0}^{M-1} y_r (q) p_1 (t-(qM+r)T_H-kT_H
\end{equation}

Introducing a new variable $n=qM+r$
\begin{equation}
g(t)=1/M \sum_{k=0}^{M-1} \sum_{n=-\infty}^\infty y(n) p_1 (t-nT_H-kT_H )
\end{equation}

The second sum above represents the conversion of $y(n)$ delayed $k$ units: $y(n-k)$. Therefore, the DT model of the architecture in Fig. 3.a represents, as shown in Fig.3.b, the signal $y(n)$ filtered by a comb filter of order $M$.
\begin{equation}
H_C (z)=1/M \sum_{k=0}^{M-1}  z^{-k} = 1/M  (1-z^{-M})/(1-z^{-k} ) 
\end{equation}

%
\section{Assessment}
The proposal is assessed in terms of robustness vs clock jitter compared with the ideal case and  previous techniques. First, the effect of the comb filter used is discussed.

The comb filter has a low-pass transfer function. Considering the oversampling ratio (OSR) as

\begin{equation}
OSR=f_H/2B
\end{equation}
\noindent where $B$ is the signal bandwidth. As long as the OSR is high and the comb-filter order, $M$, is not very large, the linear amplitude distortion produced by this filter can be tolerated.

The TI realization is based on the replacement of the classical SDM architecture by $M$ replicas clocked at a frequency $M$ times smaller. These parallel architectures are interconnected to each other by crossed paths, which increase the delay of the critical path (CP), which is the maximum-delay path that goes from the output of one register to the input of the succeeding one through combinatorial logic. The longer the CP, the lower the maximum operating frequency $f_L$. For this reason, even with highly efficient implementations, the TI modulator does not outperform the classical one for values  $M>4$.
 
The larger $M$, the smaller the -3 dB cutoff frequency of the comb filter. For $M=4$, the cutoff frequency is $f_c =0.114 f_H$. From this equation and (10), equating $f_c$ and $B$, the minimum value of the OSR that guarantees an amplitude distortion smaller than 3 dB, is 4.4. However, it is a common practice to use greater values of the OSR. 

Another effect of the comb filter, positive in this case, is the filtering of the quantization noise at high frequencies. As a matter of fact, this filter presents zeros at the frequencies $f_k = k f_H / M$, for $k = 1, 2, ..., M-1$ (for $M=4$, the zeros are at $f_1=0.25 f_H$  and $f_2=0.5 f_H$).

\subsection{Ideal case}
The architectures of Fig. 1 and Fig. 3.a are simulated with the following parameters and options. 

\begin{itemize}
    \item single-bit quantizers,
    \item  $OSR=64$,
    \item  $x(n)$ is a sinusoidal wave of frequency $f_x = B / 5$ and amplitude $A = 0.707$ (-3 dB full scale),
    \item  $M = 4$.
\end{itemize}

It is easy to show that for a second-order SDM the loop filter in Fig. 1 is $H(z)=-2z^{-1}+z^{-2}$, then the block filter results in

\begin{equation}
\bar{H}(z) = \begin{pmatrix}
 0 & -2 & 1 & 0 \\
 0 & 0 & -2 & 1 \\
 z^{-1} & 0 & 0 & -2  \\
 -2z^{-1} &  -z^{-1} & 0 & 0 \\
\end{pmatrix}
\end{equation}

The power spectrum density (PSD) of the output signals is presented in Fig. 4. It can be observed that both spectra are the same at low frequency. Note that the effect of the comb filter in the signal band is negligible (the 3 dB cut-off frequency, $f_c = 0.114$ Hz for $f_H = 1$, is higher than the signal bandwidth, $B = 0.078$ Hz). The most remarkable difference is at high frequencies, particularly, in those frequencies where the zeros of the comb filter are placed, i.e, 0.25 Hz and 0.5 Hz. Finally, the achieved signal-to-(noise+distortion) ratios (SNDR) are 69.706 and 69.715 dB for the outputs $V(w)$ and $G(w)$, respectively.

\begin{figure}[ht]
    \centering
    \includegraphics[width=15cm]{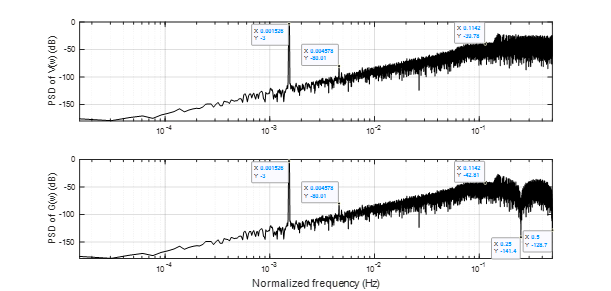}
    \caption{PSD of $V(w)$ (Fig. 1) and $G(w)$ (Fig. 3.a) from 0 to $f_H /2$.}
\end{figure}

\subsection{Effect of clock jitter}
Clock jitter refers to statistical variations in the time instants when the clock edges should ideally occur. Since these edges determine when the DAC pulse starts and ends, under clock jitter the pulse width is not constant, and the amount of signal at the DAC output randomly varies from one sampling period to the next other. The jitter can cause a severe decrease of performances of the SDM based DAC, especially at high conversion rates for wideband transmitters, where the fluctuation in the DAC pulse duration is not much smaller than the clock period. The mechanisms by which jitter produces errors are illustrated in Fig. 5 for a single-bit DAC with output levels  $\pm V_S$.

In this example, the logic sequence to be converted is the succession $0, 1, 1$ and $0$ in the sampling periods $n-1 $to $n+2$, respectively. Ideal conversion times are $t_n = n T$, where $T$ is the conversion period, which is also the ideal width of the NRZ pulses. Due to jitter, the $n$-th conversion instant is changed by the amount $\tau_n$. This produces  conversion periods to change from one to another. The jitter can have two different kinds of statistical properties. Firstly, it can be modelled as an independent noise. In this case, the time deviation $\tau_n$ is considered to be an independent and identically distributed (i.i.d.) random variable of standard deviation $\sigma_\tau$, whose power is uniformly distributed in the spectrum (white noise). The effect is that the spectral components of the out-of-band quantization noise (frequencies from $B$ to $f_s/2 $ in Fig. 4) are folded into the signal band. If the rate $\sigma_\tau/T$ is high enough, the jitter background noise can be above the in-band quantization noise and the SNDR of the SDM based DAC is degraded. 

Secondly, in systems where the clocks are generated with a voltage controlled oscillator, the so-called accumulated jitter $\tau_n$  is no longer well modelled as an i.i.d. random variable. In this case, not only the noise floor produced by the jitter is added to the in-band quantization noise, the input tone (as the one shown in Fig. 4 at the frequency 0.001526 Hz) is widened at its base and the undesirable skirts can even contain more power than the quantization noise. 

In what follows, and for the sake of simplicity, the robustness against jitter of the architectures in Fig. 1 and Fig. 3.a, will be analyzed using the the i.i.d model for clock jitter.

Under the assumption that $\tau_n$ is an i.i.d. random variable,  the signal-to-noise ratio (SNR) is given by

\begin{equation}
    SNR_{jtt,1} = \left( \frac{A}{ f_s \sigma_T \sigma_{\delta y}} \right)^2 OSR
\end{equation}
\noindent where $f_s =1/T$ is the conversion rate and $\sigma_{\delta y}$ represents the  the standard deviation of $\delta y(n)$, defined as $\delta y(n)=y(n)-y(n-1)$.


\begin{figure}[ht]
    \centering
    \includegraphics[width=7cm]{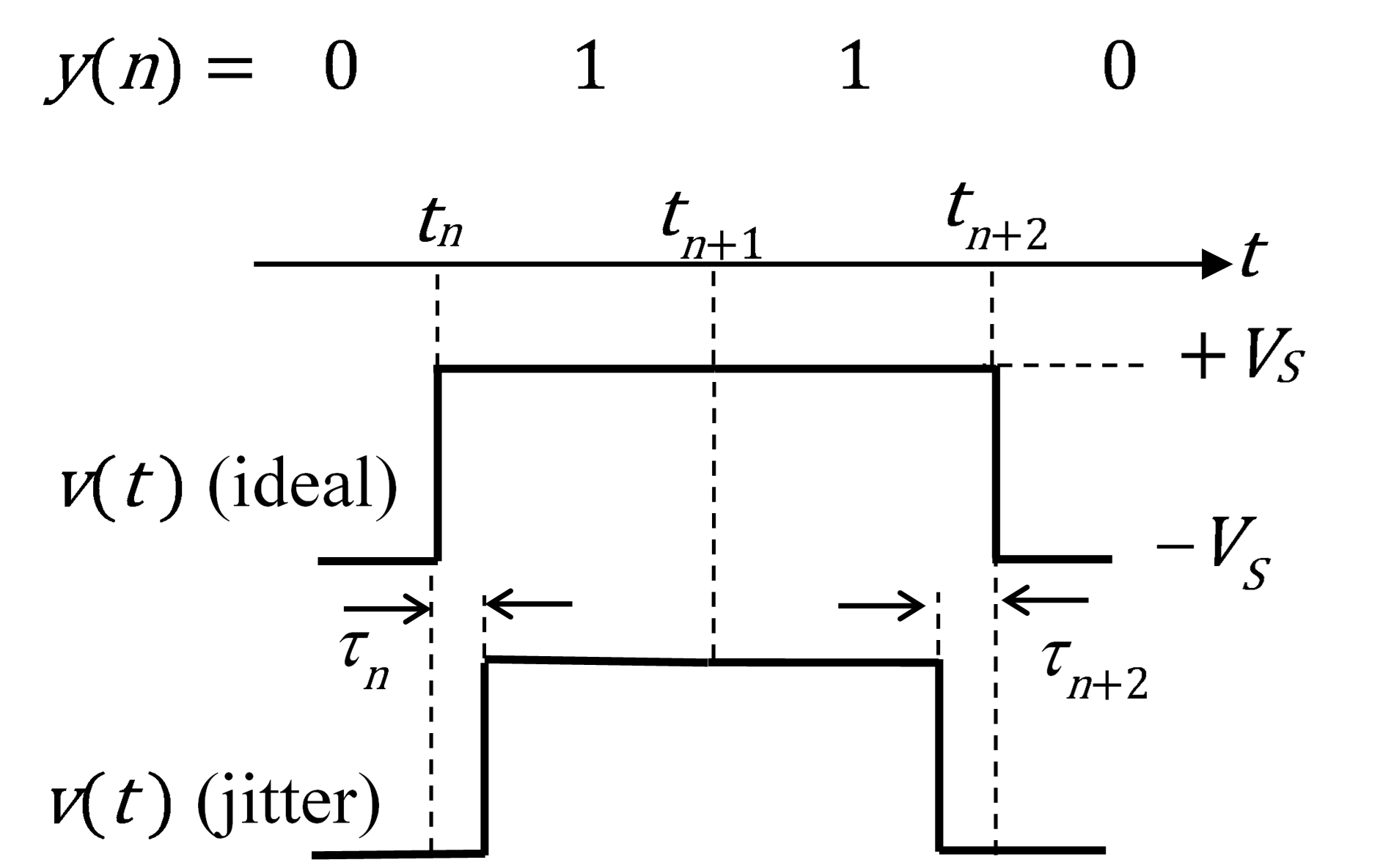}
    \caption{Effect of jitter on the waveform at the output of the DAC.}
\end{figure}

The improvement of the $SNR_{jtt}$ in the proposal, the DT model is be taken as a reference. The voltage swing at the DAC output in Fig. 3.b is assumed  to be $\pm V_S$. Although $y(n)$ is a single-bit signal, due to the comb filter, the DAC input in Fig. 3.b has $(M+1)$ levels. Therefore, the DAC step is $\Delta M$ found as

\begin{equation}
    \Delta M = \frac{2 V_S}{M} = \frac{ \Delta_1}{M}
\end{equation}

Equations (12), (13) and (14) show that the SNR of the proposed SDM based DAC ($SNR_{jtt,M}$) is $M$ 2 larger than the one in the architectures of Fig. 1 or Fig. 2 whenever the jitter is the dominant source of noise.

\begin{equation}
   SNR_{jtt,M} = M^2 SNR_{jtt,1}
\end{equation}

As a result, the proposed architecture with analog multiplexing (Fig. 3a) can achieve the same SNR as the conventional SDM based DAC (Fig. 1) even though the jitter standard deviation, $\sigma_\tau$, is $M$ times greater.

The dynamic range (DR) curve  has been obtained from simulation for architectures in Fig. 1, Fig. 2, and Fig. 3a. The same parameters of the previous example  have been chosen: $OSR=64$, normalized high rate of $f_H =1$ Hz, signal bandwidth $B = f_H/(2 OSR)$, input frequency $f_x = B / 5$ and $M = 4$. The standard deviation of the jitter has been chosen at a value high enough to be the dominant source of noise; that is, $\sigma_\tau = 0.015/fH$ seconds, which corresponds to a $SNR_{jtt,1}=47$dB for an input amplitude of $A = -3$ dB. The results are presented in Fig. 6. From (15), the expected improvement of 12 dB in the performances is in good agreement with the simulation results.

\begin{figure}[ht]
    \centering
    \includegraphics[width=14cm]{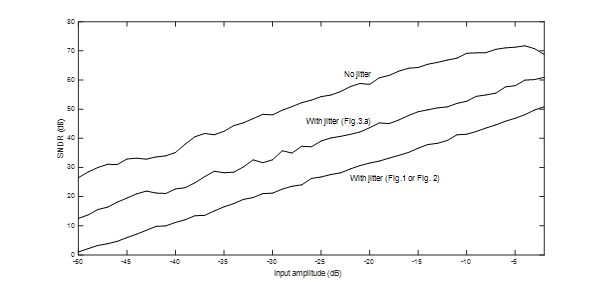}
    \caption{DR graphs of the architectures under comparison.}
\end{figure}

The spectra are  shown in Fig. 7 for an input amplitude of -3 dB. When compared to the spectra in Fig. 4, the presence of jitter is so dominant that it completely hides the quantization noise. Finally, note that the signal contaminated by jitter is no longer discrete in time. Therefore, to calculate the spectra and the SNDR with Matlab, a low-pass filtering of cut-off frequency $B$ is necessary to avoid aliasing before sampling the output signal once again. Therefore, the spectra fall at high frequencies.

\begin{figure}[ht]
    \centering
    \includegraphics[width=15cm]{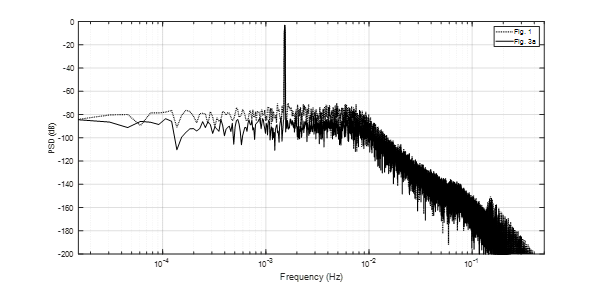}
    \caption{PSDs of $V(w)$ (Fig. 1) and $G(w)$ (Fig. 3.a) when the dominant source of noise is jitter.}
\end{figure}

%
\section{Conclusion}
The analysis of analog multiplexing for TI SDM base DACs in this preprint shows the advantages in converting to analog the $M$ low-rate signals coming from the TI SDM by means of low-speed DACs. The robustness against jitter increases in 20 log10 (M) dB or, in other words, the SDM based DAC can tolerate a jitter with a standard deviation $M$ times larger. 

In addition, the proposed architecture is more robust against the shape of the pulses of the DACs, once again in 20 log10 (M) dB. Moreover, due to the inherent advantages of RZ pulses, with time constants and SRs of only half of the clock period and 1.5 VS fH, respectively, and with a standard deviation as large as 5 \%, the SNDR does not fall respect to the ideal value of 70.9 dB, when 5 instead of 4 DACs are used for the case $M=4$.

%
\bibliographystyle{unsrt}
\bibliography{sample}

@article{li201821,
  title={A 21-GS/s single-bit second-order delta--sigma modulator for FPGAs},
  author={Li, Haolin and Breyne, Laurens and Van Kerrebrouck, Joris and Verplaetse, Michiel and Wu, Chia-Yi and Demeester, Piet and Torfs, Guy},
  journal={IEEE Transactions on Circuits and Systems II: Express Briefs},
  volume={66},
  number={3},
  pages={482--486},
  year={2018},
  publisher={IEEE}
}

@inproceedings{cordeiro2014gigasample,
  title={Gigasample time-interleaved delta-sigma modulator for FPGA-based all-digital transmitters},
  author={Cordeiro, Rui Fiel and Oliveira, Arnaldo SR and Vieira, Jose and Silva, Nelson V},
  booktitle={2014 17th Euromicro Conference on Digital System Design},
  pages={222--227},
  year={2014},
  organization={IEEE}
}

@inproceedings{tanio2017fpga,
  title={An FPGA-based all-digital transmitter with 9.6-GHz 2nd order time-interleaved delta-sigma modulation for 500-MHz bandwidth},
  author={Tanio, Masaaki and Hori, Shinichi and Tawa, Noriaki and Kunihiro, Kazuaki},
  booktitle={2017 IEEE MTT-S International Microwave Symposium (IMS)},
  pages={149--152},
  year={2017},
  organization={IEEE}
}

@article{he2025100,
  title={A 100-MHz bandwidth continuous-time sigma-delta ADC with 1 V supply in 28 nm CMOS},
  author={He, Ben and Guo, Xuan and Jia, Hanbo and Liu, Xinyu},
  journal={Microelectronics Journal},
  volume={158},
  pages={106597},
  year={2025},
  publisher={Elsevier}
}

@article{khoini1997time,
  title={Time-interleaved oversampling A/D converters: Theory and practice},
  author={Khoini-Poorfard, Ramin and Lim, Lysander B and Johns, David A},
  journal={IEEE Transactions on Circuits and Systems II: Analog and Digital Signal Processing},
  volume={44},
  number={8},
  pages={634--645},
  year={1997},
  publisher={IEEE}
}

@article{pham2008time,
  title={A Time-Interleaved Sigma Delta-DAC Architecture Clocked at the Nyquist Rate},
  author={Pham, Jennifer and Carusone, Anthony Chan},
  journal={IEEE Transactions on Circuits and Systems II: Express Briefs},
  volume={55},
  number={9},
  pages={858--862},
  year={2008},
  publisher={IEEE}
}

@article{kozak2000efficient,
  title={Efficient architectures for time-interleaved oversampling delta-sigma converters},
  author={Kozak, Mucahit and Karaman, Mustafa and Kale, Izzet},
  journal={IEEE Transactions on Circuits and Systems II: Analog and Digital Signal Processing},
  volume={47},
  number={8},
  pages={802--810},
  year={2000},
  publisher={IEEE}
}

@book{candy1991oversampling,
  title={Oversampling delta-sigma data converters: theory, design, and simulation},
  author={Candy, James C and Temes, Gabor C},
  year={1991},
  publisher={John Wiley \& Sons}
}

@inproceedings{cheng2011nyquist,
  title={Nyquist-rate time-interleaved current-steering DAC with dynamic channel matching},
  author={Cheng, Long and Ye, Fan and Yang, Hai-Feng and Li, Ning and Xu, Jun and Ren, Jun-Yan},
  booktitle={2011 IEEE International Symposium of Circuits and Systems (ISCAS)},
  pages={5--8},
  year={2011},
  organization={IEEE}
}

@article{deveugele2004parallel,
  title={Parallel-path digital-to-analog converters for Nyquist signal generation},
  author={Deveugele, Jurgen and Palmers, Pieter and Steyaert, Michiel SJ},
  journal={IEEE Journal of Solid-State Circuits},
  volume={39},
  number={7},
  pages={1073--1082},
  year={2004},
  publisher={IEEE}
}

@inproceedings{hovakimyan2019digital,
  title={Digital correction of time interleaved DAC mismatches},
  author={Hovakimyan, Garen and Hojabri, Pirooz and Martin, Gregory A and Satarzadeh, Patrick},
  booktitle={2019 IEEE International Symposium on Circuits and Systems (ISCAS)},
  pages={1--5},
  year={2019},
  organization={IEEE}
}

@article{mccue2016time,
  title={A Time-Interleaved Multimode Delta Sigma RF-DAC for Direct Digital-to-RF Synthesis},
  author={McCue, Jamin J and Dupaix, Brian and Duncan, Lucas and Mathieu, Brandon and McDonnell, Samantha and Patel, Vipul J and Quach, Tony and Khalil, Waleed},
  journal={IEEE Journal of Solid-State Circuits},
  volume={51},
  number={5},
  pages={1109--1124},
  year={2016},
  publisher={IEEE}
}

@article{colodro2003multirate,
  title={Multirate single-bit Sigma Delta modulators},
  author={Colodro, F. and Torralba, A.},
  journal={IEEE Transactions on Circuits and Systems II: Analog and Digital Signal Processing},
  volume={49},
  number={9},
  pages={629--634},
  year={2003},
  publisher={IEEE}
}

@article{colodro2010continuous,
  title={Continuous-time sigma--delta modulator with a fast tracking quantizer and reduced number of comparators},
  author={Colodro, F. and Torralba, A.},
  journal={IEEE Transactions on Circuits and Systems I: Regular Papers},
  volume={57},
  number={9},
  pages={2413--2425},
  year={2010},
  publisher={IEEE}
}

@article{colodro2014linearity,
  title={Linearity enhancement of {VCO}-based quantizers for {SD} modulators by means of a tracking loop},
  author={Colodro, F. and Torralba, A.},
  journal={IEEE Transactions on Circuits and Systems II: Express Briefs},
  volume={61},
  number={6},
  pages={383--387},
  year={2014},
  publisher={IEEE}
}

@article{colodro2022time,
  title={Time-Interleaving Sigma--Delta Modulator-Based Digital-to-Analog Converter With Time Multiplexing in the Analog Domain},
  author={Colodro, F. and Mart{\'\i}nez-Heredia, J.M. and Mora, J.L. and Ramirez-Angulo, J. and Torralba, A.},
  journal={IEEE Transactions on Circuits and Systems II: Express Briefs},
  volume={70},
  number={2},
  pages={441--445},
  year={2022},
  publisher={IEEE}
}

@article{colodro2020open,
  title={Open loop sigma-delta modulators for digital-to-analog converters with high speed improving using time interleaving},
  author={Colodro, F and Martinez-Heredia, JM and Mora, JL and Torralba, A},
  journal={AEU-International Journal of Electronics and Communications},
  volume={125},
  pages={153394},
  year={2020},
  publisher={Elsevier}
}

@article{arahal2024multi,
  title={Multi-phase weighted stator current tracking using a hyper-plane partition of the control set},
  author={Arahal, MR and Satu{\'e}, MG and Barrero, F},
  journal={Control Engineering Practice},
  volume={153},
  pages={106114},
  year={2024},
  publisher={Elsevier}
}

@article{colodro2009analog,
  title={An analog squaring technique based on asynchronous sigma--delta modulation},
  author={Colodro, F and Torralba, A and Mora, J.L. and Martinez-Heredia, J.M.},
  journal={IEEE Transactions on Circuits and Systems II: Express Briefs},
  volume={56},
  number={8},
  pages={629--633},
  year={2009},
  publisher={IEEE}
}

@article{colodro2010spectral,
  title={Spectral analysis of pulsewidth-modulated sampled signals},
  author={Colodro, F and Torralba, A},
  journal={IEEE Transactions on Circuits and Systems II: Express Briefs},
  volume={57},
  number={8},
  pages={622--626},
  year={2010},
  publisher={IEEE}
}

@inproceedings{colodro1996cellular,
  title={Cellular neuro-fuzzy networks {(CNFNs)}, a new class of cellular networks},
  author={Colodro, F and Torralba, A},
  booktitle={Proceedings of IEEE 5th International Fuzzy Systems},
  volume={1},
  pages={517--521},
  year={1996},
  organization={IEEE}
}

@article{martin2016multiphase,
  title={Multiphase rotor current observers for current predictive control: A five-phase case study},
  author={Mart{\'\i}n, C. and Arahal, M. R and Barrero, F. and Dur{\'a}n, Mario J},
  journal={Control Engineering Practice},
  volume={49},
  pages={101--111},
  year={2016},
  publisher={Elsevier}
}

@article{arahal2016harmonic,
  title={Harmonic analysis of direct digital control of voltage inverters},
  author={Arahal, M. R and Barrero, F. and Ortega, M. G and Martin, C.},
  journal={Mathematics and Computers in Simulation},
  volume={130},
  pages={155--166},
  year={2016},
  publisher={Elsevier}
}

@article{arahal2021adaptive,
  title={Adaptive cost function FCSMPC for 6-phase IMs},
  author={Arahal, M. R and Satu{\'e}, M. G and Barrero, F. and Ortega, M. G},
  journal={Energies},
  volume={14},
  number={17},
  pages={5222},
  year={2021},
  publisher={MDPI}
}

\end{document}